\documentclass[12pt,preprint]{aastex}
\usepackage{epsfig}
                                                                                
\def\vkm{km s$^{-1}$}

\def\degree{$^\circ$}
\def\arcs#1{$#1''$}
\def\arcsa#1#2{$#1^{\prime\prime}_{^\textrm{.}}#2$}

\def\solarmass{$M_\odot$}
\def\solarmasse{M_\odot}

\def\mstar#1{$M_\ast \sim #1 M_\odot$}
\def\rdisk{$r_\textrm{\scriptsize disk}$}

\def\mJyb{mJy beam$^{-1}$}
\def\Jybk{Jy beam$^{-1}$ km s$^{-1}$}
\def\mJybk{mJy beam$^{-1}$ km s$^{-1}$}

\def\cmc{cm$^{-3}$}
\def\cms{cm$^{-2}$}

\def\VLSR{V_\textrm{\scriptsize LSR}}
\def\Vsys{V_\textrm{\scriptsize sys}}
\def\Voff{V_\textrm{\scriptsize off}}
\def\rd{r_\textrm{\scriptsize D}}
                                                                                
\def\ra#1#2#3#4{#1^\mathrm{h} #2^\mathrm{m} #3^\mathrm{s}_{^\textrm{.}} #4}
\def\dec#1#2#3#4{#1\degr #2\arcmin #3^{\prime\prime}_{^\textrm{.}}#4}

\def\mHa{m_\textrm{\scriptsize H}}
\def\mH2{m_{\textrm{\scriptsize H}_2}}

\def\ro{r_\textrm{\scriptsize 0}}

\def\To{T_\textrm{\scriptsize 0}}
\def\no{n_\textrm{\scriptsize 0}}

\def\vro{v_{r\textrm{\scriptsize 0}}}
\def\vphio{v_{\phi\textrm{\scriptsize 0}}}
\def\xHCOP{$x_{\textrm{\scriptsize HCO}^{+}}$}

\def\H2{H$_2$}
\def\N2HP{N$_2$H$^+$}
\def\HCOP{HCO$^+$}
\def\cCO{C$^{18}$O}

\def\bCO{$^{13}$CO}
\def\NH3{NH$_3$}

\def\HCOP{HCO$^+$}

\def\putfig#1#2#3{\epsfig{scale=#1,angle=#2,figure=#3}}
\def\leftblank#1{}

\begin{document}

\title{ALMA Results of the Pseudodisk, Rotating disk, and Jet in Continuum
and \HCOP{} in the Protostellar System HH 212}
\author{Chin-Fei Lee\altaffilmark{1}, Naomi Hirano\altaffilmark{1},
Qizhou Zhang\altaffilmark{2}, Hsien Shang\altaffilmark{1}, 
Paul T.P. Ho\altaffilmark{1,2}, and Ruben Krasnopolsky\altaffilmark{1}
% Naomi Hirano\altaffilmark{1},
%Hsien Shang\altaffilmark{1}, and Qizhou Zhang\altaffilmark{3}
}
\altaffiltext{1}{Academia Sinica Institute of Astronomy and Astrophysics,
P.O. Box 23-141, Taipei 106, Taiwan; cflee@asiaa.sinica.edu.tw}
\altaffiltext{2}{Harvard-Smithsonian Center for Astrophysics, 60 Garden
Street, Cambridge, MA 02138}

\begin{abstract}

HH 212 is a nearby (400 pc) Class 0 protostellar system showing several
components that can be compared with theoretical models of core collapse. 
We have mapped it in 350 GHz continuum and \HCOP{} J=4-3 emission with ALMA
at up to $\sim$ \arcsa{0}{4} resolution.  A flattened envelope and a compact
disk are seen in continuum around the central source, as seen before.  The
\HCOP{} kinematics shows that the flattened envelope is infalling with small
rotation (i.e., spiraling) into the central source, and thus can be
identified as a pseudodisk in the models of magnetized core collapse.  Also,
the \HCOP{} kinematics shows that the disk is rotating and can be
rotationally supported.  In addition, to account for the missing \HCOP{}
emission at low-redshifted velocity, an extended infalling envelope is
required, with its material flowing roughly parallel to the jet axis toward
the pseudodisk.  This is expected if it is magnetized with an hourglass
B-field morphology.  We have modeled the continuum and \HCOP{} emission of
the flattened envelope and disk simultaneously.  We find that a jump in
density is required across the interface between the pseudodisk and the
disk.  A jet is seen in \HCOP{} extending out to $\sim$ 500 AU away from the
central source, with the peaks upstream of those seen before in SiO.  The
broad velocity range and high \HCOP{} abundance indicate that the \HCOP{}
emission traces internal shocks in the jet.

\end{abstract}


\section{Introduction}

Stars are formed inside molecular cloud cores by means of gravitational
collapse.  The details of the process, however, are complicated by the
presence of magnetic fields and angular momentum.  As a result, in addition
to infall (or collapse), rotation and outflow are also seen toward
star-forming regions.  
In theory, a rotationally supported disk (RSD)
is expected to form inside a collapsing core around a
protostar, from which part of the material is accreted by the protostar and
part is ejected away.  The RSD is expected to
be Keplerian when the mass of the protostar dominates that of the disk.
Observationally, however, when and how such a disk is
actually formed are still unclear, because of the lack of detailed kinematic
studies inside the collapsing core in the early phase of star formation.

%, with their radius growing with time \citep{Terebey1984}. 

RSDs have been seen with a radius of $\sim$ 500 AU in the late (i.e., Class
II or T Tauri) phase of star formation \cite[see, e.g.,][]{Simon2000}.  Such
disks must have formed early in the Class 0 phase, as claimed in a few Class
0 sources, e.g., HH 211 \cite[\mstar{0.05}, \rdisk{} $\lesssim$ 80
AU,][]{Lee2009}, NGC 1333 IRAS 4A2 \cite[\mstar{0.08}, \rdisk{} $\sim$ 130
AU,][]{Choi2010}, L 1527 \cite[\mstar{0.5}, \rdisk{} $\sim$ 125
AU,][]{Tobin2012}, and recently VLA 1623 \cite[\mstar{0.2}, \rdisk{}
$\gtrsim$ 150 AU,][]{Murillo2013}.  In models of non-magnetized core
collapse, a RSD can indeed form as early as in the Class 0 phase
\citep{Terebey1984}.  However, a realistic model should include magnetic
field, because recent survey toward a few Class 0 sources shows that
molecular cores are magnetized and likely to have an hourglass B-field
morphology \citep{Chapman2013}.  Unfortunately, in many current models of
magnetized core collapse, the magnetic field produces an efficient magnetic
braking that removes the angular momentum and thus prevents a RSD from
forming at the center \citep{Allen2003,Mellon2008}.  In those cases, only a
flattened envelope called the pseudodisk can be formed around the central
source \cite[e.g.,][]{Allen2003}.  Magnetic-field-rotation misalignment is
sometimes able to solve this so-called magnetic braking catastrophe
\citep{Joos2012,Li2013}, but not always.

This paper is a follow-up study of the HH 212 protostellar system.  This
system is deeply embedded in a compact molecular cloud core in the L1630
cloud of Orion at a distance of 400 pc.  The central source is the Class 0
protostar IRAS 05413-0104, with a bolometric luminosity
$L_\textrm{\scriptsize bol}\sim$ 9 $L_\odot$ (updated for the distance of
400 pc) \citep{Zinnecker1992}.  It drives a powerful bipolar jet
\citep{Zinnecker1998,Lee2007}.  A RSD must have formed in order to launch
the jet, accordingly to current jet models.  Previous observations in \cCO{}
($J=2-1$) and \bCO{} ($J=2-1$) with the Submillimeter Array (SMA) at $\sim$
\arcsa{2}{5} resolutions showed a flattened envelope around the central
source \citep{Lee2006}.  It is infalling with small rotation (i.e.,
spiraling) into the central source, and thus can be identified as a
pseudodisk.  The derived infall velocity and rotation velocity in the
flattened envelope suggested a small RSD to be present at the center with a
radius of $\sim$ 100 AU.  However, previous \HCOP{} J=4-3 observations
toward the inner part of the flattened envelope with the SMA at $\sim$
\arcs{1} resolution failed to confirm the presence of such a disk
\citep{Lee2007}.  Later observations at higher angular resolution in
continuum suggested again a compact disk around the central source
\citep{Codella2007,Lee2008}.  

In order to confirm the presence of a RSD in this system, we have mapped
this system in 350 GHz continuum and \HCOP{} J=4-3 using Atacama Large
Millimeter/Submillimeter Array (ALMA) at higher resolution and sensitivity. 
As before, we can identify a disk around the central source inside the
flattened envelope.  We model the continuum and \HCOP{} emission
simultaneously, deriving the kinematic and physical properties of both the
flattened envelope and the disk.  The disk indeed could have a Keplerian
rotation profile.  In addition, in our model, in order to produce enough
continuum emission of the disk, a density jump is required across the
interface between the envelope and the disk.  We will discuss how the disk
can be formed inside the envelope.  Since \HCOP{} also traces the jet close
to the central source, we will also discuss the properties and origin of the
\HCOP{} emission in the jet.

\section{Observations}\label{sec:obs}

Observations of the HH 212 protostellar system were carried out with ALMA on
2012 December 1 during the Early Science Cycle 0 phase.  Two scheduling
blocks (SBs) were obtained in Band 7 at $\sim$ 350 GHz with 21-24 antennas
in the (Cycle 0) extended configuration, with projected baselines ranging
from $\sim$ 20 to 360 m.  The correlator was set up to have 4 spectral
windows, with one for CO J=3-2, one for SiO J=8-7, one for HCO$^+$ J=4-3,
and one for the continuum.  The spectral resolution for the first 3 windows
was set to have a velocity resolution of $\sim$ 0.2 \vkm{} per channel.  A
9-pointing mosaic was used to observe the jet in this system within $\sim$
\arcs{35} from the central source.  In this paper, we only present the
observational results in \HCOP{} and continuum, discussing the envelope and
disk around the central source.  Since \HCOP{} also traces the jet close to
the central source, we will discuss the jet there too.  Observational
results in CO and SiO will be presented in next paper to discuss the jet
further out from the central source.

The data were calibrated with the CASA package, with Quasars J0538-440 (a
flux of 1.55$\pm 0.10$ Jy) and J0607-085 (a flux of 1.1$\pm 0.10$ Jy) as
passband calibrators, Quasar J0607-085 as a gain calibrator, and Callisto
and Ganymede as flux calibrators.  With super-uniform weighting, the
synthesized beam has a size of \arcsa{0}{50}$\times$\arcsa{0}{45} at a
position angle (P.A.) of 52\degree{} in continuum, and
\arcsa{0}{48}$\times$\arcsa{0}{45} at a P.A.  of 26\degree{} in HCO$^+$. 
The rms noise level is 0.61 \mJyb{} (i.e., 27 mK) for the continuum, and
$\sim$ 10 \mJyb{} (i.e., 0.45 K) for the \HCOP{} channel maps.  The
velocities in the channel maps are LSR.  The systemic velocity in this
region is assumed to be $\Vsys= 1.7\pm0.1$ \vkm{} LSR, as in \citet{Lee2007}. 
Throughout this paper, we define an offset velocity $\Voff = \VLSR - \Vsys$
to facilitate our presentation.

\section{Results}

\subsection{350 GHz Continuum}

\begin{figure} [!hbp]
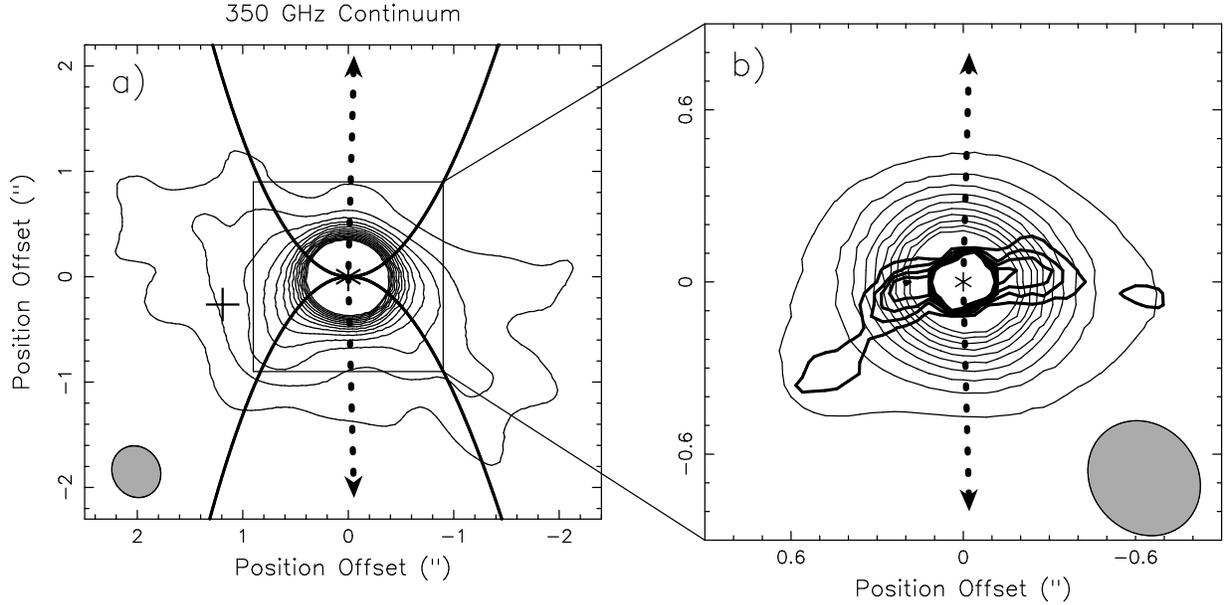

\centering
\putfig{0.65}{270}{f1.ps} % {cont_ALMA.ps}
\figcaption[]
{350 GHz continuum maps toward the central source (marked by an asterisk)
rotated by 22.5\degree{} clockwise.  The contours of the high intensity part at the center
are highly crowded and thus not shown in these maps.
The dashed arrows indicate the jet axes. 
(a) The map obtained using all available visibility data.  The synthesized
beam has a size of \arcsa{0}{50}$\times$\arcsa{0}{45} with P.A.=
52\degree{}.  The contour levels start with 5 $\sigma$ and have a step of
6 $\sigma$, where $\sigma=$ 27 mK. The peak at the center has a brightness 
temperature of $\sim$ 10 K.
The parabolic curves outline the outflow
cavity walls. The cross marks the secondary continuum peak detected in
\citet{Lee2008}.  (b) The maps obtained using visibility
data with $uv$-distance $\geq 120$ $k\lambda$.  Thin contours show the
restored map, starting with 7 $\sigma$ and having a step of 25 $\sigma$, where
$\sigma=27$ mK. The peak has a brightness temperature of $\sim$ 10 K. The synthesized beam has a size of
\arcsa{0}{42}$\times$\arcsa{0}{37} with P.A.= 63\degree{}.  Thick contours
show the CLEAN component map, starting with 0.75 K and having a step of 2.25 K.
\label{fig:cont}}
\end{figure}

The continuum emission at $\sim$ 350 GHz has been detected before with the
SMA \citep{Lee2007,Lee2008}.  It is thermal dust emission arising from a
flattened envelope perpendicular to the jet axis and a compact disk at the
center.  With high sensitivity of ALMA, the flattened envelope is seen
extending out to $\sim$ \arcs{2} (800 AU) from the central source (Figure
\ref{fig:cont}a), further away than that seen before.  It has a mean half
thickness of $\sim$ \arcs{1} (400 AU).  Its inner part, however, is shaped
by the cavity walls (outlined by the parabolic curves obtained by fitting
the CO outflow shells, Lee et al.  in prep).  The continuum peak is at
$\alpha_{(2000)}=\ra{05}{43}{51}{4107}$,
$\delta_{(2000)}=\dec{-01}{02}{53}{167}$, and it is shifted by $\sim$
\arcsa{0}{05} to the southeast as compared to that seen before.  This
shift is only a tenth of the beam size and thus could be due to a position
uncertainty in our observations.  In the following, this position is used as
the source position.  The symmetry axis (i.e., minor axis) of the
flattened envelope appears to be slightly ($<10^\circ$) misaligned with the
jet axis.  Part of this misalignment could be caused by an outflow
contamination and by a lack of emission in the northwest. In previous
observation, a secondary continuum peak was detected at $\sim$ \arcsa{1}{2}
to the southeast of the source \citep{Lee2008}.  As seen in our map, this
peak (marked as a cross) is likely to be a part of the flattened envelope in
the southeast.

\begin{figure} [!hbp]
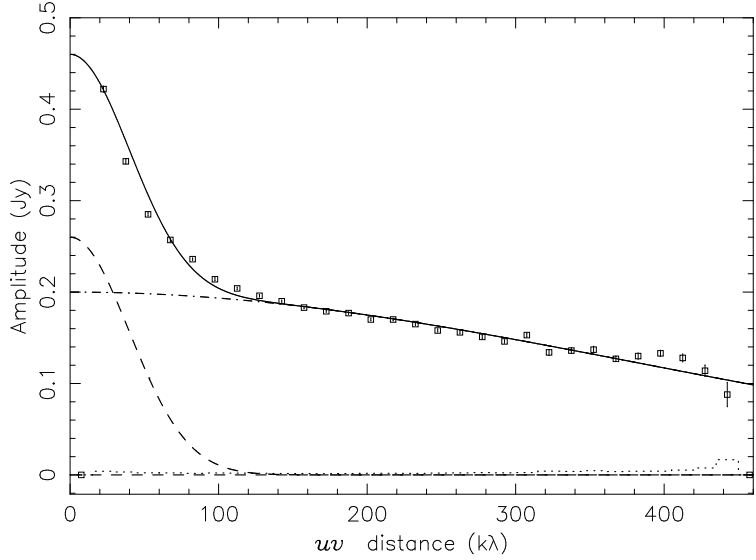

\centering
\putfig{0.4}{270}{f2.ps} % {uvamp.ps}
\figcaption[]
{ Visibility amplitude versus $uv$-distance plot 
for the continuum emission with 1 $\sigma$ error bars.
The dotted histogram is the zero-expectation level ($\sim$ 1.25 $\sigma$).
The profile can be fitted with two Gaussian components, one extended (dashed
curve) and one compact (dot-dashed curve).
The black solid curve is the sum of the two.
\label{fig:cont_uv}}
\end{figure}

Figure \ref{fig:cont_uv} shows the visibility amplitude versus $uv$-distance
plot of the continuum emission.  It can be fitted with two circular Gaussian
components, one extended tracing the flattened envelope and one compact
tracing the disk, as seen in \citet{Lee2008}.  The extended component has a
Gaussian deconvolved size of $\sim$ \arcs{2} and a flux of $\sim$ 260 mJy,
as found before.  On the other hand, the compact component has a Gaussian
deconvolved size of $\sim$ \arcsa{0}{2}, smaller than that found before,
which was $\sim$ \arcsa{0}{25}, but it has a flux of $\sim$ 200 mJy, higher
than that found before, which was $\sim$ 120 mJy.  The reason for this flux
increase is unknown.  
It could be due to a phase decorrelation at long
baselines and a poor $uv$-coverage in our previous SMA observations.  The
total flux integrated over a region within \arcsa{2}{0} from the source is
460$\pm43$ mJy, about 80 mJy higher than that found before due to this flux
increase of the compact component at the center.

The visibility data with $uv$-distance greater than 120 k$\lambda$ trace
only the compact component and thus can be used to reveal the disk structure
of the compact component.  As seen in Figure \ref{fig:cont}b, the map (thin
contours) obtained with these visibility data shows an elongated structure
perpendicular to the jet axis, with a peak brightness temperature of $\sim$
10 K at the center.  Its CLEAN component map clearly shows a disk structure
with a radius of $\sim$ \arcsa{0}{3} (120 AU).  Therefore, the compact
component at the center indeed traces a disk around the source.

\subsection{\HCOP{} J=4-3}

\HCOP{} J=4-3 emission has been detected before with the SMA at low velocity
with $|\Voff| \lesssim$ 3 \vkm{}, tracing the inner part of the flattened
envelope and outflow \citep{Lee2007}.  With the high sensitivity of ALMA, we
can map the envelope closer to the central source at higher angular and velocity
resolutions in order to check the presence of a RSD at the center, as
implied from the continuum result.  In addition, \HCOP{} emission is also
detected at high velocity with $|\Voff| \gtrsim$ 3 \vkm{}, tracing the jet
from the central source.  It is faint and thus was not detected before.

\subsubsection{Flattened Envelope, Disk, and Outflow}

\begin{figure} [!hbp]
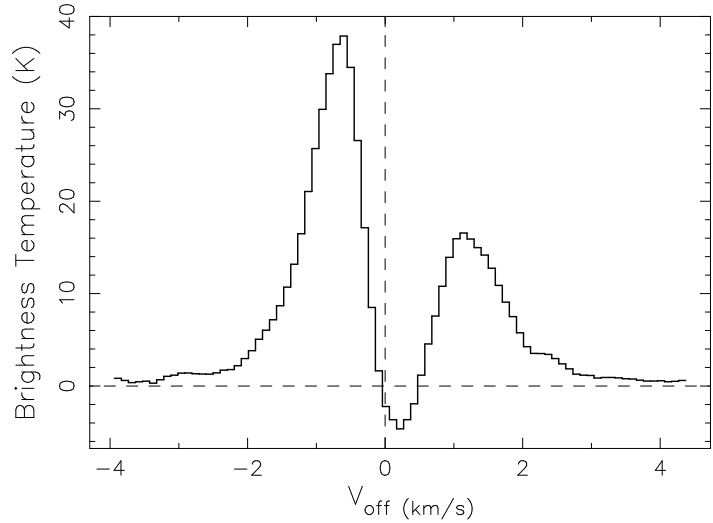

\centering
\putfig{0.75}{270}{f3.ps} %{HCOP_spec.ps}
\figcaption[]
{The \HCOP{} spectrum toward the source position, obtained by averaging over 
a rectangular region of \arcsa{1}{0}$\times$\arcsa{0}{5} oriented along the major axis
of the envelope. The dashed vertical line marks the systemic velocity. The dashed horizontal
line marks the zero brightness temperature.
\label{fig:HCOP_spec}}
\end{figure}

The \HCOP{} spectrum toward the central region
(\arcsa{1}{0}$\times$\arcsa{0}{5} oriented along the major axis) of the
envelope shows that the \HCOP{} emission is detected with $|\Voff{}|
\lesssim 3$ \vkm{} in the envelope (Figure \ref{fig:HCOP_spec}).  The
spectrum has two peaks, one in the blue and one in the red, with the one in
the blue brighter than the one in the red.  It also has a dip in between at
$\Voff{} \sim 0.2$ \vkm{} on the redshifted side.  These two spectral
features suggest an infall motion in the envelope \citep{Evans1999}.  In
order to see the velocity structure in detail, we divide the velocity into 3
ranges: $|\Voff| \leq 1$ \vkm{}, $1\leq |\Voff| \leq 2$ \vkm{}, and $2\leq
|\Voff| \leq 3$ \vkm{}.

The \HCOP{} emission with $|\Voff| \leq 1$ \vkm{} is shown in Figure
\ref{fig:HCOP}a.  In the equator, the emission traces the flattened
envelope.  The blueshifted emission is seen across the central source with a
peak to the west.  The redshifted emission is much weaker and is seen mainly
in the east.  This spatial distribution has been seen before in \bCO{} J=2-1
in \citet{Lee2006}, and it indicates that the envelope mainly has infall
motion, and some small rotation with the redshifted side in the east and
blueshifted side in the west.  The redshifted emission is fainter due to a
self-absorption in the near side of the envelope \citep{Evans1999} and it is
shifted to the east due to rotation motion.  Away from the equator, the
emission in the south shows a parabolic shell tracing the outflow cavity
wall opening to the south, with the blueshifted emission to the west and
redshifted emission to the east of the jet axis.  The velocity sense is the
same as that of the envelope, suggesting that the cavity wall consists of
swept-up material originally in the envelope.  In the north, the emission
structure is complicated by bow shock interactions.  Redshifted emission is
seen around bow shocks NF and NK1, and faint blueshifted emission is seen
around bow shock NF.  These emissions could trace the bow shock wings
interacting with the outflow cavity wall outlined by the parabolic curve. 
Also, a bright blueshifted jet-like structure (labeled as BJ) is seen
extending from the inner blueshifted part of the envelope to the northeast. 
It could trace a low-velocity part of a jet or a jet interaction with the
outflow cavity wall.

The \HCOP{} emission with $1\leq |\Voff| \leq 2$ \vkm{} is shown in Figure
\ref{fig:HCOP}b, with a zoom-in shown in Figure \ref{fig:HCOP}c.  In the
equator, the redshifted and blueshifted emissions are seen with their peaks
on the opposite sides within $\sim$ \arcsa{0}{3} of the central source
(Figure \ref{fig:HCOP}c), spatially coincident with the disk-like continuum
emission seen in Figure \ref{fig:cont}b.  The redshifted peak is shifted
slightly to the south due to a contamination of a redshifted jet-like
emission extending to the south.  Therefore, that part of the envelope
indeed has become a rotating disk around the central source.  Also, jet-like
emission starts to show up clearly, extending to the north and south out
from the disk.  Thus, the spectrum shown in Figure \ref{fig:HCOP_spec} also
includes some of these emissions at the base.  In the south, the parabolic
outflow shell is still seen, tracing the outflow cavity wall at higher
velocity.  In the north, redshifted emission is seen mainly around bow shock
NF, tracing the bow shock wings.  However, no blueshifted emission is seen
around that bow shock.  In previous study, CO emission was detected around
that bow shock not only in redshifted velocity but also in blueshifted
velocity extending to $\Voff \sim$ $-$5 \vkm{} \cite[see Fig.  5
in][]{Lee2007}, indicating that the blueshifted emission has a wider spread
of velocities, and thus lower intensity per velocity.  This is because that
bow shock is tilted toward us, so that its back wall is projected on the low
redshifted side, while its front wall is projected to a wide range of
blueshifted velocities \citep{Lee2001}.  Thus, the blueshifted \HCOP{}
emission is likely lost in the noise.

At higher velocity with $2\leq |\Voff| \leq 3$ \vkm{}, the redshifted and
blueshifted emission in the equator are also seen on the opposite sides of
the central source but closer in (Figures \ref{fig:HCOP}d and
\ref{fig:HCOP}e for a zoom-in), tracing the disk emission.  Jet emission is
clearly seen extending out to the north and south, and it seems to connect
to the disk emission.  At this high velocity, neither shell-like emission
nor emission around the bow shocks are detected.

Figure \ref{fig:pvHCOP} shows the PV diagram of the \HCOP{} emission cut
along the equator in order to further study the kinematics of the envelope
and disk.  The blueshifted peak is brighter than the redshifted peak and the
emission is missing at low redshifted velocity ($\Voff \sim$ 0 to 0.5
\vkm{}), consistent with that seen in the spectrum in Figure
\ref{fig:HCOP_spec}.  As mentioned earlier, these are the well-known
signatures for an infall motion in the envelope \citep{Evans1999}, and the
missing of the redshifted emission is due to a self absorption in the near
side of the envelope.  At low velocity, a triangular PV structure is seen on
the blueshifted side with its base extending from the west to the east
across the central position and its tip pointing toward the high blueshifted
velocity, also as expected for an infall envelope \citep{Ohashi1997}.  A
similar triangular PV structure could also be seen on the redshifted side,
although its base is almost gone due to the self-absorption.  Therefore, the
low-velocity part indeed traces the infalling envelope.  Since the
triangular PV structure on the blueshifted side is shifted slightly to the
west and the triangular PV structure on the redshifted side is shifted
slightly to the east, the infalling envelope also has a small rotation.
As discussed later, the lack of emission in the low redshifted velocity is
enhanced by a presence of an extended infalling envelope that is resolved
out by the interferometer.  The negative contours centered at $\sim$ 0.2
\vkm{} around the source are due to an absorption of the bright continuum
(disk) emission near the source by the near side of the envelope.  At
$|\Voff| \gtrsim 1$ \vkm{}, the emission shrinks toward to the center, with
the blueshifted emission mostly shifted to the west and the redshifted
emission mostly shifted to the east.  This is because the rotation starts to
dominate and the innermost part of the envelope has become a rotating disk. 
The PV structure there seems to be better described by a Keplerian rotation
curve than the rotation curve with conservation of specific angular
momentum.

\subsubsection{Extended infalling envelope}

Figure \ref{fig:pvallen}a shows the PV diagram cut along the jet axis.  The
PV diagram is complicated, consisting of several components.  The faint
emission with $|\Voff| \gtrsim 2$ \vkm{} mainly traces the jet.  The bright
emission at low velocity (with $|\Voff| \lesssim 2$ \vkm) located within
$\sim$ $\pm$\arcs{1} from the central source should mostly trace the
flattened envelope.  The low-velocity emission that extends further out
($\sim$ \arcs{5} to the north and $\sim$ \arcs{3} to the south) from the
central source should trace the outflow shells and bow shock wings.  The
emission is also missing at low velocity, with the mean velocity of the
missing part indicated by the thick solid line.  In the north, the mean
velocity of the missing part is always on low redshifted side decreasing
from $\Voff \sim$ 0.2 to 0.05 \vkm{}, while in the south, the mean velocity
of the missing part goes from low redshifted velocity at $\Voff \sim$ 0.2
\vkm{} around the source to near the systemic velocity at $\sim$
\arcsa{1}{5} and beyond.  This suggests a presence of an extended infalling
envelope in the outer part of the system and its near side is projected at
those velocities, absorbing the emission from the jet, flattened envelope,
outflow shells, and bow shock wings, and then itself is resolved out by the
interferometer.

In order to have those projected velocities, the material in the extended
envelope is unlikely to be infalling radially toward the center, in which
case, it would produce similar projected velocities in both the north and
south.  Since hourglass B-field morphology has been predicted in the
extended infalling envelope in theoretical model of core collapse
\citep{Allen2003} and has also been claimed in a few Class 0 systems
\citep{Chapman2013}, this B-field morphology is expected to be present in
our system as well.  Therefore, one likely possibility to produce those
projected velocities is to have the material there flowing along the
hourglass field lines toward the flattened envelope, as seen in theoretical
model of core collapse \citep{Allen2003}.  Since the jet is tilted at a
small inclination angle with the north side toward us, the field lines are
likely tilted toward us in the north while tilted to the plane of the sky in
the south, as shown in Figure \ref{fig:pvallen}b.  Therefore, the near side
of the extended envelope is seen almost with redshifted velocity in the
north, while first seen in redshifted velocity and then seen near the systemic
velocity in the south, producing the missing part in the PV diagram.

\subsubsection{Jet}

Figure \ref{fig:HCOPjet} shows the \HCOP{} jet and its PV diagram.  Here the
map of the jet is obtained using the emission with $|\Voff| \geq 2$ \vkm{}. 
This is because the emission with $|\Voff| \leq 2$ \vkm{} is contaminated by
the emission of the flattened envelope, disk, outflow cavity walls, and bow
shock wings, as shown in Figure \ref{fig:HCOP}.  The \HCOP{} jet is
collimated.  Its redshifted emission extends mainly to the south with a peak
at $\sim$ $-$\arcsa{1}{0}, and its blueshifted emission extends mainly to
the north with a peak at $\sim$ \arcsa{0}{6}.  Its emission peaks are
located upstream of those seen before in SiO emission, which are at $\sim$
\arcs{1} in the north and \arcs{2} in the south \citep{Lee2008}.  The PV
diagram is the same one in Figure \ref{fig:pvHCOP}, but is focused on the jet
and thus shown with a wider velocity range and a smaller contour step.  It
shows that the \HCOP{} emission of the jet has a broad range of velocities,
indicating that it arises from internal shocks in the jet, like the SiO
emission \citep{Lee2008}.  The northern part of the jet emission extends
from $\sim$ $-$10 to 3 \vkm{}, and the southern part from $\sim$ $-$5 to 10
\vkm{}.  Thus, the \HCOP{} emission in the jet has a mean velocity range of
$\sim$ 14 \vkm{}.  The middle velocities of the jet emission are $\Voff
\sim$ $-$3.5 \vkm{} in the north and 2.5 \vkm{} in the south, and thus much
lower than the jet velocity \cite[which is 100$-$200 \vkm,][]{Lee2008}. 
This is consistent with the jet lying close to the plane of the sky.

\section{Model of the envelope and disk}

Here we construct a spatio-kinematic model to reproduce simultaneously the
continuum and \HCOP{} emission of the flattened envelope and disk.  Figure
\ref{fig:modelfig} shows the schematic diagram of the model in Cartesian
coordinate system.  The flattened envelope is in the equatorial plane ($x-y$
plane) perpendicular to the jet axis ($z-$axis) and the compact disk is
located at the center inside the flattened envelope.  Two outflow cavities
(obtained by fitting the CO outflows, Lee et al.  in prep) are also
included, excavating the innermost part of the flattened envelope.  However,
no extended infalling envelope is included due to a lack of physical
information.  For simplicity, dust and \HCOP{} gas are assumed to have the
same temperature.  The flattened envelope and disk are dusty and gaseous,
producing both the continuum and \HCOP{} emissions.

Previously, the outer part of the flattened envelope was detected in
C$^{18}$O J=2-1 with an outer radius of $\sim$ \arcs{5} (2000 AU)
\citep{Lee2006}.  Thus, the outer radius of the flattened envelope $r_E$ is
assumed to have this value.  The inner radius of the flattened envelope is
set to the outer radius of the disk.  The half thickness of the outer part
of the flattened envelope is found to be $\sim$ \arcsa{1}{2} in C$^{18}$O
J=2-1.  From the \HCOP{} and continuum maps, the half thickness of the inner
part of the flattened envelope is estimated to be $\sim$ \arcsa{0}{8}.  Thus
in this model, the half thickness of the flattened envelope ($H_E$) is
assumed to increase from \arcsa{0}{8} from the inner radius to \arcsa{1}{2}
at the outer radius.  The innermost part of the flattened envelope is
excavated by the outflow.

In the flattened envelope, the temperature, number density of molecular hydrogen,
infall velocity, and rotation velocity are assumed to follow those obtained
from modeling the outer part of the flattened envelope in C$^{18}$O, which are
respectively
\begin{equation}
T^{e} = \To^e (\frac{r}{\ro})^{-0.4}
\end{equation}
\begin{equation}
n^{e} = \no^e (\frac{r}{\ro})^{-1.5}
\end{equation}
\begin{equation}
v_{r}^{e} = \vro^e (\frac{r}{\ro})^{-0.5}
\end{equation}
and
\begin{equation}
v_{\phi}^{e} = \vphio^e (\frac{r}{\ro})^{-1.0}
\end{equation}
with $r$ being the radial distance from the central source.
Here $\ro=$\arcs{1} (400 AU) is the reference radius, and $\To^e$, $\no^e$,
$\vro^e$, and $\vphio^e$ are the reference values at that radius.  The
density has a power-law index of $-1.5$, as predicted in infalling models
\cite[e.g.,][]{Nakamura2000}.  The temperature has a power-law index of
$-0.4$, as found in the envelopes around other low-mass protostars.  The
flattened envelope has a radial infall velocity assumed to be that of free fall.
Its rotation velocity is assumed to be that of constant specific angular momentum. 
Thus, the material in the flattened envelope is spiraling inward toward the central source.

The disk is not well resolved in our observations, especially in the minor
axis.  It is assumed to be a flat disk with an outer radius of
$r_D=$\arcsa{0}{3} (120 AU) and a constant half thickness of $H=$\arcsa{0}{1}
(40 AU), as judged from the Clean Component map.  Note that this outer
radius of the disk can be considered as an upper limit of the actual disk
radius.  The inner radius of the disk is unknown and set to a very small
value of \arcsa{0}{01} (4 AU).  When the flattened envelope transforms into
the disk, the rotation velocity becomes Keplerian, as predicted in core
collapse model \cite[e.g.,][]{Nakamura2000} and as seen in the older system
HH 111 \citep{Lee2010}.  Thus, the rotation velocity in the disk is assumed
to be Keplerian.  With the free-fall velocity of the flattened envelope, we
can derive the mass of the central protostar to be
\begin{equation}
M_\ast = \frac{r (v_r^e)^2}{2 G} = \frac{\ro (\vro^e)^2}{2 G}
\label{eq:mastr}
\end{equation}
and then the rotation velocity of the disk to be
\begin{equation}
v_{\phi}^{D} = \sqrt{\frac{G M_\ast}{r}} = \frac{\vro^e}{\sqrt{2}} (\frac{r}{\ro})^{-0.5}
\end{equation}

In the current models of disk formation \cite[e.g.,][]{Krasnopolsky2002},
there will be an accretion shock across the envelope-disk interface,
producing a jump in the temperature, density, and infall velocity there.  As
discussed later, since the disk density is high there with a value $>$
$10^8$ \cmc{}, the cooling is efficient and thus the shock should be close
to be isothermal.  Therefore the temperature is assumed to be continuous
across the interface.  In addition, the temperature in the disk is assumed
to increase with the decreasing radius with a power-law index $q$, i.e.,
\begin{equation}
T^{D} = \To^E (\frac{r}{\rd})^{-q}
\end{equation}
where $\To^E=\To^e(\rd/\ro)^{-0.4}$ is the envelope temperature at the disk radius. 

The number density is discontinuous with a jump across the interface. The jump is
assumed to be a free parameter $m$ to be obtained from our model.
Therefore, the number density is assumed to be given by 
\begin{equation}
n^D= m \cdot \no^E (\frac{r}{\rd})^{-1} 
\end{equation} 
where $\no^E=\no^e(\rd/\ro)^{-1.5}$ is the envelope number density at the
disk radius. The density of the disk is assumed to have a power-law index
of $-1$. With a constant thickness, the disk has a surface density with a power-law
index of $-1$, similar to that found in the Class I disk \citep{Lee2011} and
T-Tauri disks \citep{Andrews2009}.

The infall velocity also has a jump across the interface. 
According to jump condition for mass conservation,
the infall velocity at the interface drops to 
\begin{equation}
\vro^D= \frac{\vro^E}{m}
\end{equation}
where $\vro^E=\vro^e(\rd/\ro)^{-0.5}$ is the envelope infall velocity
at the interface. In our model, the accretion velocity
in the disk at the interface is assumed to be given by this infall velocity.
The accretion velocity in the disk, $v_r^D$, can then be derived,
assuming a constant accretion rate across the disk, i.e.,
$v_r^D n^D H r$  = constant.
Since $n^D\propto r^{-1}$ and $H$=constant, the accretion velocity is constant and given by
\begin{equation}
v_r^D= \vro^D = \textrm{constant}
\label{eq:accvel}
\end{equation}

A similar but more self-consistent model derived from the theoretical work
of \citet{Ulrich1976} has been applied to massive star-forming
regions \citep{Keto2010}.  Our model here is simple and can be considered as
a first step to disentangle the envelope and disk in our system.  In our model
calculations, radiative transfer is used to calculate the continuum and
\HCOP{} emissions, with an assumption of local thermal equilibrium.  To
calculate the continuum emission, we assume a dust mass opacity law given by
\cite[see, e.g.,][]{Beckwith1990} 
\begin{equation} 
\kappa_\nu = 0.1
(\frac{\nu}{10^{12}\; \textrm{Hz}})^\beta \;\textrm{cm}^2 \;\textrm{g}^{-1}
\end{equation}
with $\beta= 1$, as found in \citet{Lee2007} from fitting the SED. To
calculate the \HCOP{} emission, the abundance of \HCOP{} relative to
molecular hydrogen, \xHCOP{}, is required and assumed to be a free parameter
to be derived by fitting the continuum and the \HCOP{} emission
simultaneously.  The line width of the \HCOP{} emission is assumed to be
given by the thermal line width only.  The line width due to turbulence is
not included in our model. 
We first calculate the channel maps with radiative transfer and next map them with the observed $uv$-coverage and
velocity resolutions.  The resulting channel maps can then be used to make
the continuum map and \HCOP{} channel maps. The \HCOP{} channel maps
can then be used to make the spectrum, integrated maps, and PV
diagrams of \HCOP{}.

In our model, there are 7 free parameters, with their best-fit values of
$\To^e \sim 45$ K, $\no^e \sim 5\times10^6$ \cmc{}, $\vro^e \sim 0.9$
\vkm{}, $\vphio^e \sim 0.35$ \vkm{}, $m \sim 8.0$, $q \sim 0.6$, and \xHCOP
$\sim 10^{-9}$.  The distributions of the rotation velocity, infall
velocity, sound speed, temperature, and density of the envelope and disk
derived from these parameters are shown in Figure \ref{fig:fitfunc}.  The
best-fit parameters are obtained by matching the continuum map, \HCOP{}
channel maps, spectrum, and PV diagram in our model to those in the
observations (Figure \ref{fig:modelfit}).  As expected, the best-fit
parameters in the envelope are similar to those obtained from modeling the
outer part of the envelope seen in \cCO{} \citep{Lee2006}.  The \HCOP{} abundance is also
similar to that found in molecular cloud cores, which is $(0.5-2) \times
10^{-9}$ \citep{Girart2000}.  The temperature power-law index of the disk is
also similar to that found in the Class I disk in HH 111 \citep{Lee2011}. 
Notice that since the disk is not well resolved, this power-law index here
should be considered as a rough value.  The high density jump in the disk is
required to produce the bright disk emission at the center.  Without the
density jump, the continuum flux at the center would be a factor of $\sim$ 7
lower than the observed.  This density jump will be further discussed later.

Our model can reproduce the continuum and \HCOP{} emission of the flattened
envelope and disk reasonably well.  As can seen in Figure
\ref{fig:modelfit}a, our model can reproduce the continuum structure and
intensity for both the flattened envelope and compact disk.  Our model is
symmetric and thus is not intended to produce the emission extending only to
the south.  Our model can also reproduce the PV diagram of the \HCOP{}
emission along the major axis (Figure \ref{fig:modelfit}e).  Note that the
low-velocity emission with $|\Voff| \lesssim 0.5$ \vkm{} is expected to
mostly disappear if an extended infalling envelope is included.  To support
our argument, we assume that the effect of the extended infalling envelope
toward the center mimics a Gaussian spectrum centered at $\sim$ 0.2 \vkm{}. 
After we subtract this Gaussian component from our model spectrum (red
spectrum), we can obtain a spectrum (green spectrum) that matches the
observed spectrum toward the center at low velocity (Figure
\ref{fig:modelfit}d).  Our model can also reproduce the \HCOP{} maps at $1
\leq |\Voff| \leq$ 2 \vkm{} (Figures \ref{fig:modelfit}b and c).  Since our
model does not include the outflow and jet, it can not reproduce their
emission seen in the observations.  In the \HCOP{} spectrum, part of the
observed redshifted emission with $\Voff \gtrsim$ 1 \vkm{} is from them, and
thus can not be reproduced in our model.

\subsection{Consistency Checks}

Here in this section, we present consistency checks for the best-fit
parameters in our model.  Keplerian radius is the radius where the rotation
velocity equals the Keplerian rotation.  Thus, equaling the Keplerian rotation
velocity to the rotation velocity in the flattened envelope, we can derive
the Keplerian radius to be
\begin{equation} 
r_K = 2 \ro (\frac{\vphio^e}{\vro^e})^{2} 
\end{equation}
With the best-fit parameters, the Keplerian radius is found to be $\sim$
\arcsa{0}{3}, which turns out to be the same as the outer radius of the disk
assumed here in our model.  Therefore, the disk could indeed have an outer
radius of $\sim$ \arcsa{0}{3} (120 AU).  Note that since the disk is not
spatially resolved in our observations, this disk radius should be
considered as an upper limit.

As discussed earlier, an accretion shock is expected to be present
at the interface between
the envelope and disk, producing a jump across the interface.  
At low angular resolution, a compact SO emission with a radius of
$\lesssim 120$ AU was detected 
around the central source of HH 212, with a low-velocity component
tracing the rotation \citep{Lee2007}. Since
SO emission has been claimed to trace accretion shocks in the disks of other sources,
e.g., HH 111 \citep{Lee2010} and L1527 \citep{Sakai2014}, the SO emission 
in HH 212 could
trace an accretion shock as well. Further observations at higher resolution
are needed to confirm this.
For an isothermal shock, the density is expected to have
a jump factor of $m = M^2$, where $M$ is the shock Mach number.  The
isothermal shock Mach number at the disk radius can be given by
\begin{equation}
M \approx \frac{|\vro^E-\vro^D|}{c_s} = \frac{m-1}{m}\frac{|\vro^E|}{c_s}
\end{equation}
where $c_s$ is the isothermal sound speed of the envelope at the disk
radius.  At the disk radius, the temperature of the envelope is $\sim$ 73 K
and thus the isothermal sound speed is $c_s = \sqrt{\frac{kT}{\mu
\mHa}}\sim$ 0.51 \vkm{}, here $\mu=2.33$ as Helium is included with
$n_\textrm{\scriptsize He}= 0.1 n_\textrm{\scriptsize H}$.  The infall
velocity of the envelope at the disk radius is $\vro^E=-1.64$ \vkm{} and thus
$M\sim 2.82$.  Therefore, $m\sim 8.0$, the same as that obtained from our
model, and thus the jump condition is self-consistent in our model. 

Initially, the shock should be adiabatic. According to shock jump
condititons with $M \sim 2.82$, the gas should be compressed and heated by
the shock to have a density of $\sim 10^8$ \cmc{} and a temperature of
$\sim$ 200 K.  Since the density is high, the cooling due to dust grains is
efficient.  Using the cooling rate of dust grains in \citet{Suttner1997},
the cooling time is estimated to be $\sim$ 1 yr.  With the accretion
velocity of $\sim$ 0.2 \vkm{}, the material only moves inward by a distance
of $\sim$ 0.04 AU, much smaller than the disk size.  Thus, the shock should
quickly become radiative and isothermal, consistent with our model.

Using Eq. \ref{eq:mastr}, the mass of the central star is found to be $\sim$
0.18 \solarmass{}.  The disk mass is 
\begin{equation} M_D = 1.4\mH2
\int^{r_D} 2 \pi r (2 H) n^D dr \sim 0.014 \; \solarmasse 
\end{equation} 
We can check this disk mass using the observed continuum flux of the disk. 
Assuming that the observed disk continuum emission is optically thin, this
mass requires the disk to have a representative temperature of $\sim$ 91 K. 
In our model, this temperature corresponds to the value at $r \sim$
\arcsa{0}{2} in the disk.  In the observations, from the SED fitting, the
representative temperature of the envelope and disk together was found to be
$\sim$ 48 K \citep{Lee2007}.  The disk is warmer and thus can have a
representative temperature as high as 91 K.
As a result, the disk mass derived here should be consistent with the observations.
It is $\sim$ 8\% of the stellar mass and thus the disk indeed can be formed. 
On the other hand, the mass in the flattened envelope is
\begin{equation} 
M_E = 1.4\mH2
\int^{r_E}_{r_D} 2 \pi r (2 H_E) n^e dr \sim 0.10 \; \solarmasse 
\end{equation}
which is $\sim$ 56\% of the stellar mass.

In our model, the accretion velocity in the disk is 
assumed to be given by the infall velocity at the shock interface
(see Eq. \ref{eq:accvel}). It is $\sim$ 0.20
\vkm{} and thus much smaller than the rotation velocity (Figure \ref{fig:fitfunc}a),
as expected for a RSD.  The accretion rate in the disk is the
same as the infall rate at the disk radius, and thus it is given by
\begin{equation} \dot{M}_D = 1.4 \mH2 2 \pi r_D (2H) \no^E \vro^E \sim
5.0\times 10^{-6} \; \solarmasse{}\; \textrm{yr}^{-1} \end{equation}
Assuming the same accretion rate in the past, the accretion time would be
$0.18/5.0\times 10^{-6} \sim 3.6\times10^4$ yrs, as expected for a Class 0
source.  The accretion luminosity is $L_\textrm{acc}= G
M_\ast\dot{M}/R_\ast$.  Assuming $R_\ast \sim 2 R_\odot$, then
$L_\textrm{acc} \sim 14$ L$_\odot$, comparable to the bolometric luminosity
of this source, which is $\sim$ 9 L$_\odot$ \cite[][corrected for the new
distance of 400 pc]{Zinnecker1992}.

\section{Discussion}

\subsection{Comparing to Class 0 and I disks}

Our model is a simple spatio-kinematic model, consisting of a flattened
envelope with its material spiraling toward the central source and a RSD at
the center.  This model is inspired by the previous work on the Class I
system HH 111 in Orion in the later stage of star formation \citep{Lee2010}. 
In that system, the rotation velocity is found to change from that of
conserving angular momentum in the flattened envelope to that of Keplerian
in the disk.  Here we assume the same change of rotation velocity here in HH
212.  Judging from the Clean Component map of the continuum, the disk can
have a radius of $\sim$ 120 AU (\arcsa{0}{3}).  This disk radius turns out
to be the same as the Keplerian radius derived from the infall velocity and
rotation velocity profiles found in our model.  This disk radius is also
similar to those claimed in other Class 0 systems
\citep{Lee2009,Choi2010,Tobin2012,Murillo2013}.  This radius is much smaller
than that in HH 111, which is $\sim$ 2000 AU.  This is likely because the
disk size grows with time \citep{Terebey1984}.  The accretion age is $\sim$
$4\times10^5$ yrs in HH 111 \citep{Lee2010}, a factor of $\sim$ 10 older
than that in HH 212.  In addition, the specific angular momentum in the
infalling flattened envelope is $\sim$ 1550 AU \vkm{} in HH 111, also a
factor of $\sim$ 10 larger than that in HH 212, which is
$l=\ro\cdot\vphio^e$ $\sim$ 140 AU \vkm{}.  Therefore, in HH 111, the
expansion wave could have expanded $\sim$ 10 times further away, bringing in
more material with higher specific angular momentum from a much larger
distance \citep{Terebey1984}.  A few other Class I Keplerian disks have been
recently claimed in Taurus in \citet{Harsono2013}.  Those disks have a disk
radius of $\sim$ 100 AU, similar to that found in the Class 0 disks.  The
small radius of those disks could be due to the low specific angular
momentum of $\lesssim$ 200 AU \vkm{} in their infalling envelopes.  In
another word, the disk radius could depend significantly on the
environmental conditions.

In our model of HH 212, there is a huge jump (a factor of $\sim$ 8) in density
and infall (radial) velocity across the interface between the envelope and
disk in the equatorial plane.  This is because the infall velocity is much
higher than the isothermal sound speed at the interface, producing a
(accretion) shock Mach number $M\sim 3$.  In HH 111, on the other hand, the
infall velocity is only slightly higher than the isothermal sound speed at
the interface, resulting in a small Mach number $M<1.4$.  Thus, the jump in
density and infall velocity is not significant.  In another word, accretion
shock and thus density jump would be less obvious in the later phase of star
formation as the disk grows larger, because infall velocity decreases faster
with the increasing distance from the source than the sound speed.

% 1 km/s /pc = 3.2320880E-14 s^-1

The formation of Class 0 disk is still unclear in theory because some of the
current models of magnetized core collapse include significant magnetic
braking \citep{Krasnopolsky2002,Allen2003,Mellon2008}, large enough to
hinder disk formation unless somehow overcome through the effects of (e.g.)
turbulence, asymmetry, or non-ideal MHD.  In HH 212, the formation of the
disk could be facilitated by the outflow walls at the base, which force the
material to flow in a small channel toward the central protostar.  Further
observations are needed to check this possibility.

%This could slow down the flow and help the formation of a disk.  

\subsection{Pseudodisk and Extended Envelope}

The flattened envelope is infalling with rotation (i.e., spiraling) into the
central source, and thus can be identified as a pseudodisk found in the
models of magnetized core collapse \cite[see, e.g.,][]{Allen2003}.  The
outer radius of this pseudodisk could be $\sim$ 2000 AU, judging from
previous \cCO{} observations \citep{Lee2006}.  Using the mean isothermal
sound speed in the pseudodisk, which is $\sim$ 0.4 \vkm{} (see Figure
\ref{fig:fitfunc}a), the dynamical age of the pseudodisk could be $\sim$
$2.4\times10^4$ yrs.  This age is about two-third of the accretion time
estimated earlier, acceptable in the models of magnetized core collapse
\cite[see, e.g.,][]{Allen2003}.  Since the pseudodisk is essentially a
magnetic feature, polarization observations with ALMA in thermal dust
emission are needed to confirm the presence of magnetic field in the
pseudodisk and check if the field axis is aligned with the the symmetry
axis of the pseudodisk.  As mentioned earlier, the symmetry axis of the
flattened envelope appears to be slightly ($<10^\circ$) misaligned with the
jet axis.  Further work is needed to study if this misalignment could be due
to a magnetic-field-rotation misalignment that could help the formation of
the disk \citep{Joos2012}.

In addition, an extended infalling envelope is required to be present
surrounding the pseudodisk in order to produce the missing flux in the PV
diagram (Fig.  \ref{fig:pvallen}) and the spectrum in \HCOP{} (Fig. 
\ref{fig:HCOP_spec}) at low-redshifted velocity.  This extended envelope,
however, is resolved out by ALMA in our observations.  In this extended
envelope, the material is required to be flowing roughly parallel to the jet
axis toward the pseudodisk.  This is expected if the envelope is magnetized
with an hourglass B-field morphology roughly parallel to the jet axis. 
Since the material can flow along the field lines, the material can flow
parallel to the jet axis into the pseudodisk.  In order to confirm this
scenario, future polarization observations with ALMA (with Compact Array) in
thermal dust emission are really needed to map the B-field morphology in the
extended envelope.

\subsection{\HCOP{} Jet and Abundance}

The jet is clearly seen in \HCOP{}.  The \HCOP{} abundance there could be
highly enhanced in order for the \HCOP{} emission to be detected.  To
explore this possibility, here we derive the \HCOP{} abundance using the
constraint on the mass-loss rate to be derived from the \HCOP{} emission. 
Since the jet emission with $|\Voff| \leq 2$ \vkm{} is partly missing and is
partly contaminated by the outflow/envelope/disk emission, the real flux of
the \HCOP{} emission in the jet can be estimated by first measuring the flux
with $|\Voff| \geq 2$ \vkm{}, and then multiplied it by 14/10 = 1.4, where
14 \vkm{} is the mean velocity range of the jet and 10 \vkm{} is the
velocity range used in the measurement of the flux.  The \HCOP{} jet using
$|\Voff| \geq$ 2 \vkm{} has a mean flux of $\sim$ 0.7 \Jybk{} along the jet
axis.  Thus, the corrected mean flux is $\sim$ 1.0 \Jybk{}.  The excitation
temperature of the \HCOP{} emission of the jet is assumed to be 50 K, like
that of the CO emission of the jet \citep{Lee2007,Cabrit2012}.  The \HCOP{}
emission of the jet has a brightness temperature of only 2-10 K, and thus
can be assumed to be optically thin.  Therefore, the mean column density of
\HCOP{} is $N \sim 2.0 \times 10^{13} $ \cms{}.  Note that an increase of 50
K in the excitation temperature only produces an increase of $\sim$ 30\% in
the column density.

The mean density in the jet can be given by
\begin{equation}
n_j = \frac{N}{x d_j} \frac{b_\perp}{d_j}
\label{eq:nj}
\end{equation} 
where $b_\perp \sim$ \arcsa{0}{5} is the beam size across the jet axis as
the jet is not resolved and $d_j$ is the jet diameter.
Thus, the mass-loss rate in two sides of the jet can then be given by
\begin{equation} 
\dot{M}_j = 2 \frac{\pi d_j^2}{4} n_j v_j (1.4 \mH2) = \frac{\pi}{2} \frac{N}{x} b_\perp v_j (1.4 \mH2)
       \sim \frac{1 \times10^{-13}}{x} \;\;\solarmasse{}\; \textrm{yr}^{-1}
\end{equation}
where $v_j \sim 150$ \vkm{} is the jet velocity \citep{Lee2008}, and $x$
is the \HCOP{} abundance to be derived.  In the current MHD jet launching
models, the mass-loss rate can have a mean of 20\% of the accretion
rate \citep{Shu2000,Konigl2000}.  Therefore, with the accretion rate derived
earlier, we have $\dot{M}_j \sim 1 \times 10^{-6}$ \solarmass{}
yr$^{-1}$.  As a result, we have $x \sim 1\times 10^{-7}$, higher than
that derived in the flattened envelope by a factor of $\sim$ 100. 
This abundance is also consistent
with that found in other sources, in which the \HCOP{} abundance has been
greatly enhanced in shocks with a range from $1\times10^{-8}$ to
$7\times10^{-7}$ \citep{Viti2002,Dent2003,Rawlings2004}.

We now can derive the mean density in the jet using Equation \ref{eq:nj} and
the \HCOP{} abundance.  The jet was found to have a diameter of $d_j\sim$
\arcsa{0}{2} (80 AU) \citep{Cabrit2007,Lee2008}, thus the mean density in
the jet is $\sim 4.2 \times 10^5$ \cmc{}.  For the two \HCOP{} emission
peaks in the jet, their fluxes are $\sim$ 60\% higher than the mean flux of
the jet, and thus their density can be 60\% higher, or $\sim 6.7 \times
10^5$ \cmc{}.  This density is more than 10 times lower than the critical
density of \HCOP{} J=4-3 line, which is $\sim 9 \times 10^6$ \cmc{}.

The two \HCOP{} peaks in the jet are seen with a broad velocity range of
$\sim$ 14 \vkm{}, suggesting that they are produced by internal shocks of
$\sim$ 14 \vkm{} in the jet.  Previous observations show that SiO emissions
are seen downstream produced by stronger shocks of $\sim$ 20 \vkm{}
\citep{Lee2008}.  That no \HCOP{} emission is seen downstream suggests that
\HCOP{} is either destroyed by stronger shocks or excited to higher J levels
in a warmer environment.  In any case, the shock seems to become stronger
with the increasing distance from the source, as expected if the shock is
produced by a variation in jet velocity \citep{Raga1990}.  The peak
positions are asymmetric with the northern one at $\sim$ \arcsa{0}{6} (220
AU) and the southern one at $\sim$ \arcsa{1}{0} (400 AU).  Assuming a jet
velocity of 150 \vkm{}, the dynamical ages are $\sim$ 8 and 13 yrs,
respectively.  Thus, the abundance enhancement can take place in less than
$\sim$ 10 yrs.  The process for this enhancement is unclear.  It is possibly
the result of ice mantle evaporation and/or sputtering and may also be due
to enhanced gas-phase production in dense and warm environments
\citep{Rawlings2013}.  Further study is really needed to determine the
process that can lead to this abundance enhancement in the internal shocks
of the jet.

\section{Conclusions}

We have mapped the HH 212 protostellar system in the dust continuum at 350
GHz and the \HCOP{} (J=4-3) line.  Our primary conclusions are the
following:

\begin{itemize}

\item A flattened envelope is seen in continuum and \HCOP{} around the
central source, extending out to $\sim$ 800 AU (\arcs{2}).  The \HCOP{}
kinematics shows that the flattened envelope is infalling with rotation
(i.e., spiraling) into the central source, and thus can be identified as a
pseudodisk found in the models of magnetized core collapse.

\item  Inside the flattened envelope, a bright compact disk is seen in
continuum at the center with an outer radius of $\sim$ 120 AU
(\arcsa{0}{3}).  This disk is also seen in \HCOP{} and the \HCOP{}
kinematics shows that the disk is rotating.

\item \HCOP{} emission is missing at low-redshifted velocity centered at
$\Voff \sim$ 0.2 \vkm{}.  To account for this missing, an extended infalling
envelope is required, with its material flowing roughly parallel to the jet
axis toward the pseudodisk.  This is expected if the envelope is magnetized
with an hourglass B-field morphology.  Since the material can flow along the
field lines and the field lines are roughly parallel to the jet axis, the
material can flow parallel to the jet axis into the pseudodisk.

\item We have modeled the continuum and \HCOP{} emission of the flattened
envelope and disk simultaneously.  In our model, we assume a change of
rotation profile from that of conserving specific angular momentum in the
flattened envelope to that of Keplerian in the disk, as inspired by
theoretical models of core collapse and previous observations of HH 111.  In
order to reproduce enough disk continuum emission at the center, a jump in
density with a factor of $\sim$ 8 is required across the interface between
the flattened envelope and the disk.  This jump turns out to be consistent
with an isothermal shock at the interface.  In our model, the disk only has
$\sim$ 8\% of the stellar mass and thus can indeed be formed.  However,
further observations at higher resolution are needed to confirm its
Keplerian rotation and its radius.

\item A collimated jet is seen in \HCOP{} extending out to $\sim$ 500 AU
from the central source (and disk?), with its peaks located upstream of
those seen before in SiO.  The \HCOP{} emission is seen with a broad range
of velocities.  The \HCOP{} abundance is highly enhanced by a factor of
$\sim$ 100, comparing to that of the flattened envelope.  All these suggest
that the \HCOP{} emission traces internal shocks in the jet.

\end{itemize}

\acknowledgements

This paper makes use of the following ALMA data:
ADS/JAO.ALMA\#2011.0.00647.S.  ALMA is a partnership of ESO (representing
its member states), NSF (USA) and NINS (Japan), together with NRC (Canada)
and NSC and ASIAA (Taiwan), in cooperation with the Republic of Chile.  The
Joint ALMA Observatory is operated by ESO, AUI/NRAO and NAOJ.  These data
were made available to Chin-Fei Lee, as part of his ALMA proposal
2011.0.00122.S (PI: Chin-Fei Lee), which requested observations duplicating
those of proposal 2011.0.00647.S.  C.-F.  Lee acknowledges grants from the
National Science Council of Taiwan (NSC 101-2119-M-001-002-MY3) and the
Academia Sinica (Career Development Award).

%These data were made available to Chin-Fei Lee  [*] as part of
%his/her/their ALMA proposal 2011.0.00122.S (PI: Chin-Fei Lee) [**],
%which requested observations duplicating those of proposal 2011.0.00647.S."

%[*] The author(s) who got access to the considered data may plausibly be
%one or several Co-Is of the considered proposal; the PI may not
%necessarily be involved in the publication.

%[**] If the PI is the only author appearing in the statement, this
%parenthesis can be omitted.  On the other hand, it may be better to give
%the full PI name rather than just the last name to avoid any ambiguity.

%% Remember to include "(" and ")" for the year,e.g., (1998)
%%

\begin{figure} [!hbp]
\centering
\putfig{0.8}{270}{f4.ps} %{cont_HCOP.ps}
\figcaption[]
{
Contour maps of blueshifted and redshifted \HCOP{} emission superimposed
on the \H2{} jet (gray image) from \citet{McCaughrean2002}.  Three bow
shocks, SK1, NF, and NK1 are seen in the \H2{} jet.  Label BJ in (a) indicates a
blueshifted jet-like structure at low velocity.
Asterisk, parabolic
curves, and dashed arrows indicate the source position, outflow cavity
walls, and jet axes, respectively.  For the \HCOP{} maps, the synthesized
beam has a size of \arcsa{0}{48}$\times$\arcsa{0}{45} with P.A.=
29\degree{}.  Three velocity ranges are shown, with (a) $|\Voff| \leq 1$
\vkm{}, (b-c) $1\leq |\Voff| \leq 2$ \vkm{}, and (d-e) $2\leq |\Voff| \leq 3$
\vkm{}.  In (a)-(d), the contours start at 60 \mJybk{} and have a step of
120 \mJybk{}.  In (e), smaller contour step is used to show the detailed
structure of the \HCOP{} jet.  The contours start at 60 \mJybk{} and have a
step of 40 \mJybk{}.
\label{fig:HCOP}
}
\end{figure}

\begin{figure} [!hbp]
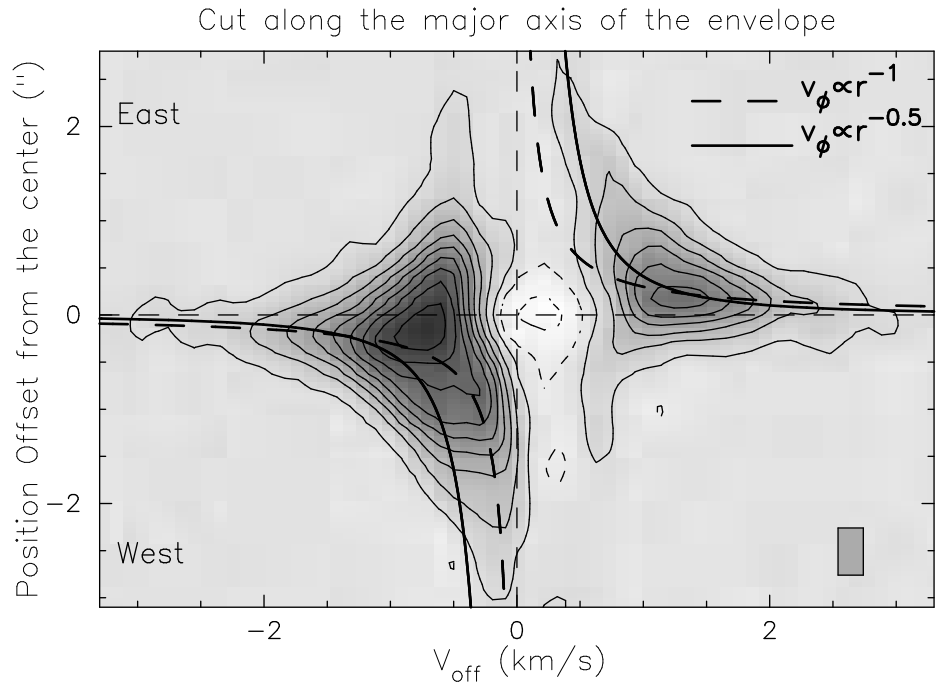

\centering
\putfig{0.7}{0}{f5.ps} %{pvHCOP_env.ps}
\figcaption[]
{Position-velocity diagram in \HCOP{} cut along the major axis of the flattened
envelope. The contours start at 1.8 K, with a step size of 5.4 K.
The solid and dashed lines show the possible
Keplerian rotation curve and the rotation curve that
conserves the specific angular momentum, respectively.
\label{fig:pvHCOP}
}
\end{figure}

\begin{figure} [!hbp]
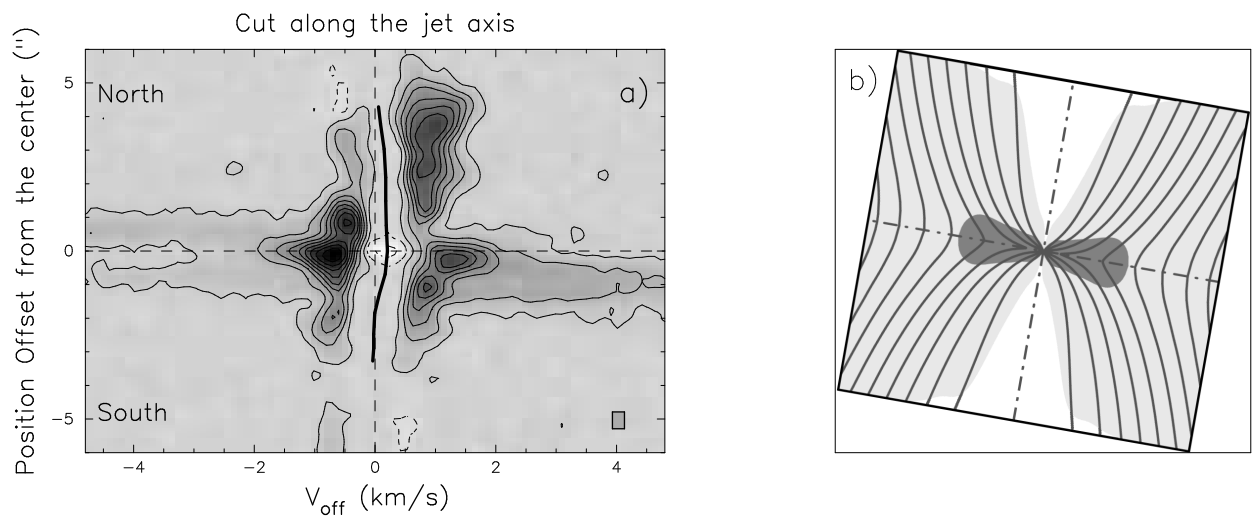

\centering
\putfig{0.7}{270}{f6.ps} %{pvHCOP_allenFig.ps} 
\figcaption[]
{(a) PV diagram in \HCOP{} cut along the jet axis.
The contours start at 1.8 K, with a step size of 5.4 K. The thick solid line
indicates the mean velocity of the missing part of the emission.
(b) Morphology of magnetic field lines predicted from a core collapse model in
\citet{Allen2003}. This figure is adopted from \citet{Chapman2013}. 
The dark gray region shows the flattened envelope
(pseudodisk) and the light gray region shows the extended infalling
envelope. To be compared with our target source, the model is rotated by
10\degree{} clockwise, so that the field lines in the extended envelope
is tilted toward us in the north while titled to near the plane of the sky
in the south, causing the missing flux in the PV diagram. 
\label{fig:pvallen}
}
\end{figure}

\begin{figure} [!hbp]
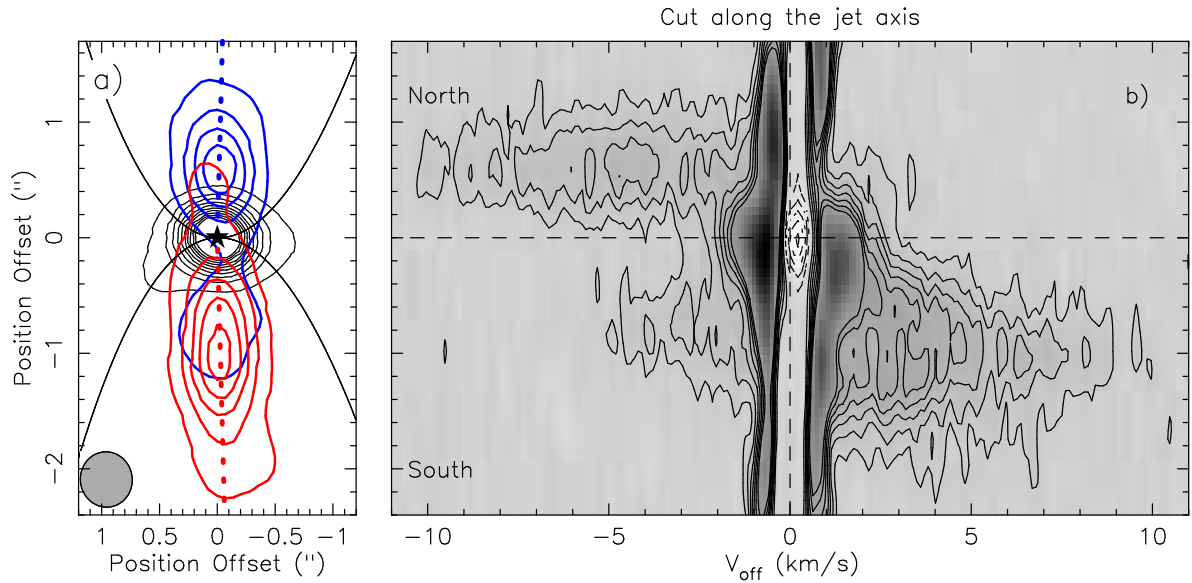

\centering
\putfig{0.65}{270}{f7.ps} %{pvHCOPjet.ps}
\figcaption[]
{(a) Contour map of \HCOP{} jet emission superimposed
on the continuum map of the disk as shown in Figure \ref{fig:cont}b.
Blueshifted and redshifted \HCOP{} emission are integrated from
$\Voff=-10$ to $-2$ \vkm{} and $\Voff=2$ to 9 \vkm{}, respectively.
The contours start at 100 \mJybk{} and have a
step of 200 \mJybk{}. (b) PV diagram of the jet emission cut along the jet
axis. The contours start at 1.8 K and have a step of 1.8 K.
\label{fig:HCOPjet}
}
\end{figure}

\begin{figure} [!hbp]
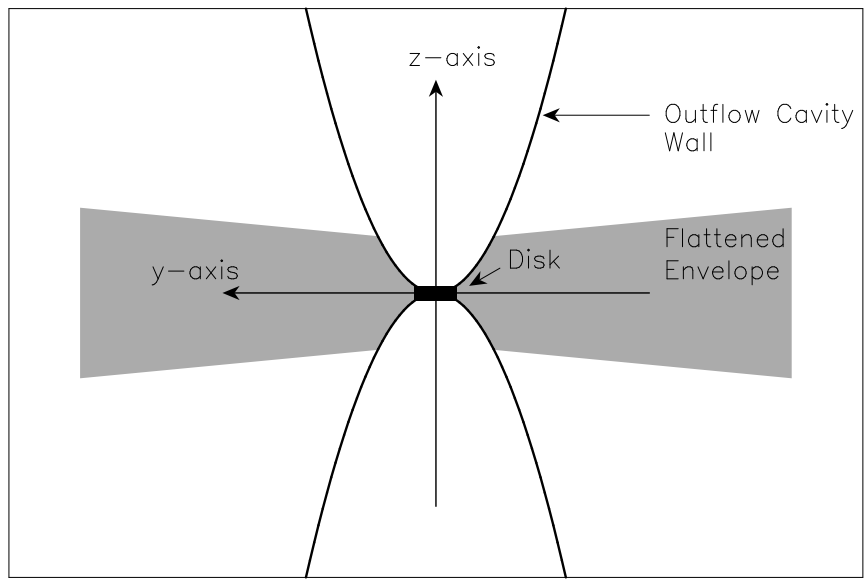

\centering
\putfig{0.5}{270}{f8.ps} %{skematic.ps}
\figcaption[]
{A schematic diagram of our model for the flattened envelope and the disk.
Jet axis is aligned with the z-axis.
\label{fig:modelfig}
}
\end{figure}

\begin{figure} [!hbp]
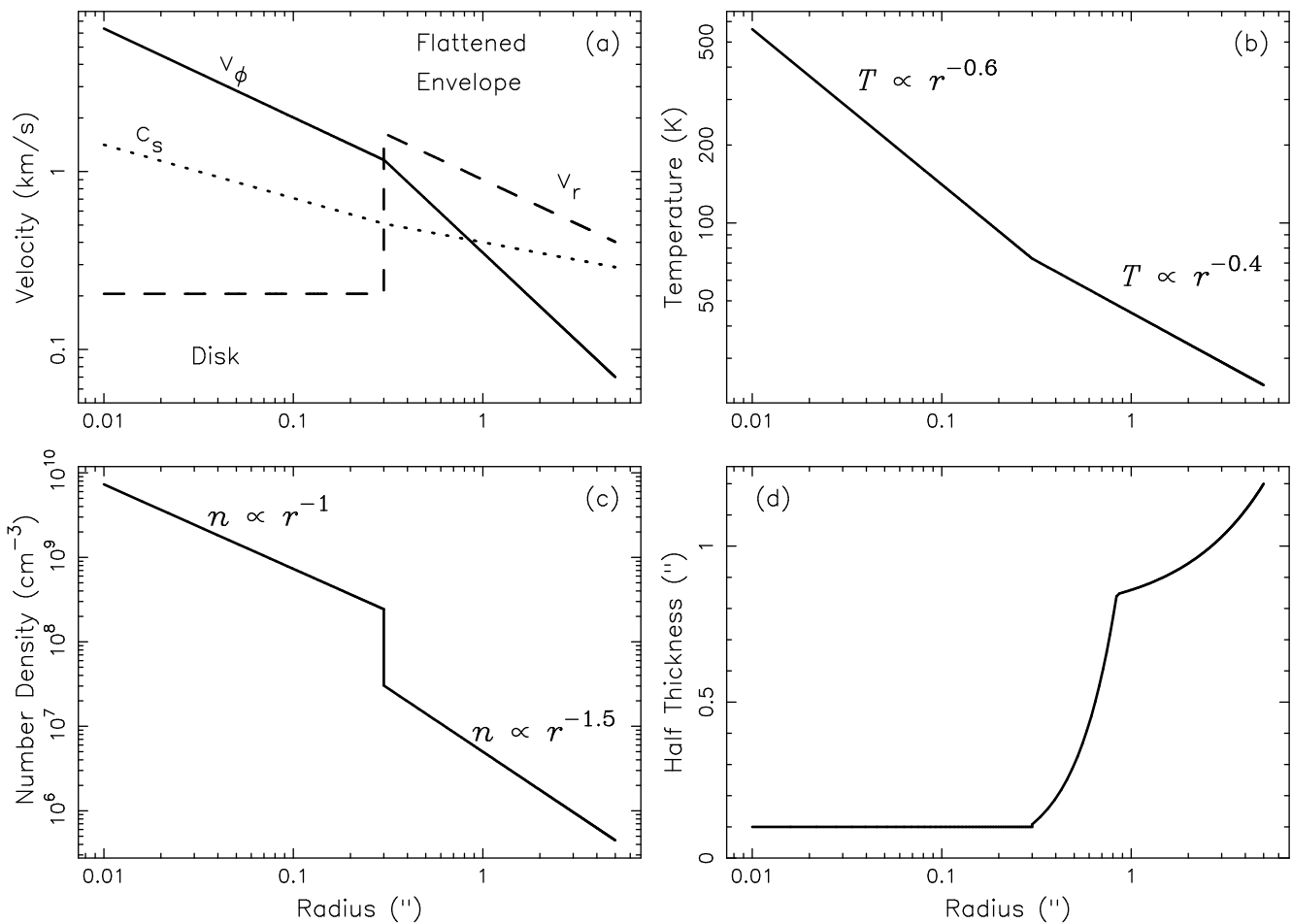

\centering
\putfig{0.7}{270}{f9.ps} %{fitfunc.ps}
\figcaption[]
{(a-c) show the distributions of the rotation velocity, infall
velocity, sound speed, temperature, and density of the flattened envelope and disk
derived from the best-fit parameters in our model.
(d) shows the half thickness assumed in our model as described in the text.
\label{fig:fitfunc}
}
\end{figure}

\begin{figure} [!hbp]
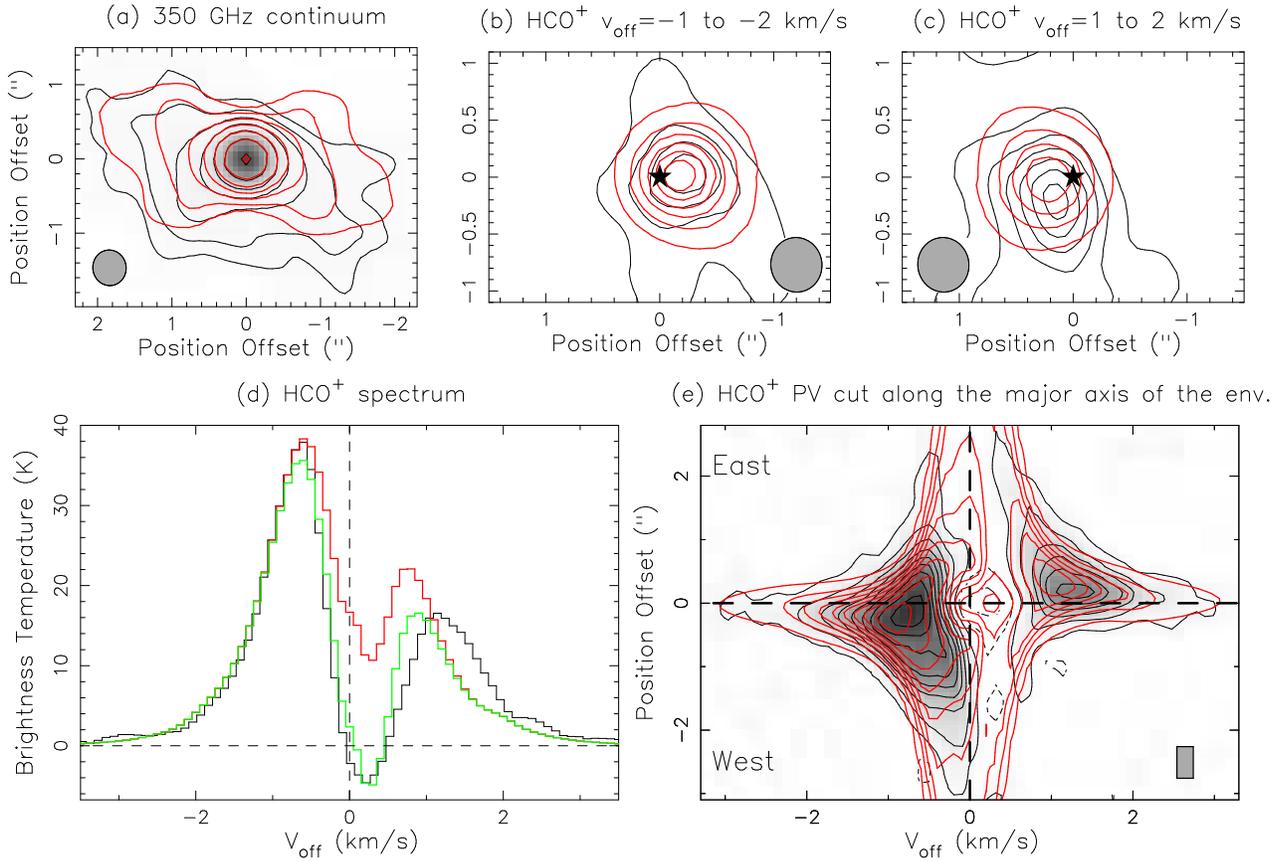

\centering
\putfig{0.65}{270}{f10.ps} %{modeln.ps}
\figcaption[]
{Our model results (red contours and spectrum) 
in comparison to the observations (black contours and spectrum). (a) shows the continuum
as shown in Figure \ref{fig:cont}a but with logarithmic contour levels to show the bright
emission at the center. The contour levels are 116 $\cdot2^{n-1}$ mK, where $n=1,2,3,..$.
(b) and (c) show the blueshifted and redshifted
\HCOP{} emission with $1 \leq \Voff \leq 2$ \vkm{}, as shown in Figure \ref{fig:HCOP}c.
(d) shows the spectrum as shown in Figure \ref{fig:HCOP_spec}.
The green spectrum  is the model spectrum (red spectrum) minus a Gaussian spectrum
centered at 0.2 \vkm{} that mimics the effect of an extended infalling
envelope. (e) shows the PV diagram as shown in Figure \ref{fig:pvHCOP}.
\label{fig:modelfit}
}
\end{figure}

\end{document}